
\input jytex.tex   
\typesize=10pt
\magnification=1200
\baselineskip=17truept
\hsize=6truein\vsize=8.5truein
\sectionnumstyle{blank}
\chapternumstyle{blank}
\chapternum=1
\sectionnum=1
\pagenum=0

\def\begintitle{\pagenumstyle{blank}\parindent=0pt\begin{narrow}[0.4in]}
\def\endtitle{\end{narrow}\newpage\pagenumstyle{arabic}}


\def\beginexercise{\vskip 20truept\parindent=0pt\begin{narrow}[10 truept]}
\def\endexercise{\vskip 10truept\end{narrow}}


\def\eql#1{\eqno\eqnlabel{#1}}
\def\ref{\reference}
\def\peq{\puteqn}
\def\pref{\putref}

\def\mgn{\marginnote}
\def\bex{\begin{exercise}}
\def\eex{\end{exercise}}


\def\mbox#1{{\leavevmode\hbox{#1}}}
\def\hspace#1{{\phantom{\mbox#1}}}


\def\be{\beta}

\def\ka{\kappa}
\def\la{\lambda}

\def\om{\omega}

\def\si{\sigma}

\def\ze{\zeta}

\def\De{\Delta}


\def\frac#1/#2{\leavevmode\kern.1em
\raise.5ex\hbox{\the\scriptfont0 #1}\kern-.1em/\kern-.15em
\lower.25ex\hbox{\the\scriptfont0 #2}}
\def\sfrac#1/#2{\leavevmode\kern.1em
\raise.5ex\hbox{\the\scriptscriptfont0 #1}\kern-.1em/\kern-.15em
\lower.25ex\hbox{\the\scriptscriptfont0 #2}}

\def\gtorder{\mathrel{\raise.3ex\hbox{$>$}\mkern-14mu
             \lower0.6ex\hbox{$\sim$}}}
\def\ltorder{\mathrel{\raise.3ex\hbox{$<$}|mkern-14mu
             \lower0.6ex\hbox{\sim$}}}

\def\semidirprod{\rlap{\ss C}\raise1pt\hbox{$\mkern.75mu\times$}}
\def\for{\lower6pt\hbox{$\Big|$}}
\def\fish{\kern-.25em{\phantom{abcde}\over \phantom{abcde}}\kern-.25em}


\def\boxit#1{\vbox{\hrule\hbox{\vrule\kern3pt
        \vbox{\kern3pt#1\kern3pt}\kern3pt\vrule}\hrule}}
\def\dalemb#1#2{{\vbox{\hrule height .#2pt
        \hbox{\vrule width.#2pt height#1pt \kern#1pt
                \vrule width.#2pt}
        \hrule height.#2pt}}}


\def\noin{\noindent}


\def\ie{{\it i.e. }}

\def\pa{\partial}


\def\tr{{\rm tr\,}}

\def\3j#1#2#3#4#5#6{\left\lgroup\matrix{#1&#2&#3\cr#4&#5&#6\cr}
\right\rgroup}

\def\man{{\cal M}}

\def\m?{\mgn{?}}


\def\cqg#1#2#3{{\it Class. Quant. Grav.} {\bf {#1}} (19{#2}) #3}

\def\jmp#1#2#3{{\it J. Math. Phys.} {\bf {#1}} (19{#2}) #3}


\begin{title}
\vglue 20truept
\righttext {MUTP/94/--}
\righttext{}
\leftline{\today}
\vskip 100truept
\centertext {\Bigfonts \bf Functional determinants}
\vskip 5truept
\centertext{\Bigfonts\bf and effective actions:}
\vskip 5truept
\centertext{\Bigfonts\bf Corrigenda}
\vskip 15truept
\centertext{J.S.Dowker\footnote{Dowker@v2.ph.man.ac.uk} and J.S.Apps}
\vskip 7truept
\centertext{\it Department of Theoretical Physics,\\
The University of Manchester, Manchester, England.}
\vskip 60truept
\centertext {Abstract}
\begin{narrow}
Some calculational errors in expressions derived previously by the first
author for the effective action, or equivalently for the functional
determinant, on sectors of a spherical cap are corrected. The formula for
the change in the effective action under Weyl rescalings in the three
dimensional case is also amended.
\end{narrow}
\vskip 5truept
\righttext {October 1994}
\vskip 75truept
\righttext{Typeset in \jyTeX}
\vfil
\end{title}
\pagenum=0
Recalculation and internal checking has revealed errors in some recent work by
the first author [\pref{Dow1,Dow2}] which we would like to rectify here.

In [\pref{Dow1}] the behaviour of the effective action under conformal
transformations was considered. The three dimensional expression quoted in
equation (26) is incorrect due to a transcription error and a minor algebraic
mistake. The formula should be
$$ \eqalign{
W_{R}[\la^2 g]-W_{R}[g]={1\over1536\pi}\int_{\pa\man}\bigg[\big(6\tr(\ka^2)
-&3\ka^2-16{\widehat\De}_2\om-4\widehat R\big)\om\cr
&+30\ka N+18N^2-24n^\mu n^\nu \om_{\mu\nu}\bigg],\cr}
\eql{26}$$
where the notation is explained in [\pref{Dow1}]. Various equivalent forms can
be found upon partial integration and the introduction of the embedding
curvature and Laplacian. Equation (\peq{26}) satisfies the required symmetry
under interchange of $g$ and $\bar g,\,=\la^2 g$.

The error in reference [\pref{Dow2}] concerning the functional determinant on a
two-dimensional spherical cap is more serious in its knock-on consequences.
It stems from a transcription error in equation (14) where $4/(1+a^2)^2$ in
the final two brackets should be $2/(1+a^2)$, this being $\sqrt h$.

The correct expression simplifies considerably to
$$
W^D[\bar g,g]={1\over6}\ln2-{2\over3}{a^2\over1+a^2}=
{1\over6}\ln2-{1\over3}\si.
\eql{14}$$
Then, using the the disc effective action derived by
Weisberger in terms of that for the hemisphere, one finds
$$
W^D_{\rm cap}(\si)=W^D_{\rm hemisphere}+{1\over3}(\si-1)-
{1\over12}\ln\big({\si\over2-\si}\big).
\eql{15}$$
This equation, with
$$
W^D_{\rm hemisphere}=-\ze_R'(-1)+{1\over8}-2\ze_R'(0),
$$
replaces (15) of [\pref{Dow2}].

The Neumann result is
$$
W^N[\bar g,g]={1\over6}\ln2+{1\over3}{a^2\over1+a^2}-{1\over2}
\ln\big({4\over1+a^2}\big)
\eql{24}$$leading to
$$
W^N_{\rm cap}(\si)=W^N_{\rm hemisphere}-{1\over6}(\si-1)+
{1\over12}\ln(2-\si)+{5\over12}\ln\si,
\eql{16}$$ with
$$
W^N_{\rm hemisphere}=-\ze_R'(-1)+{1\over8}+2\ze_R'(0).
$$
Equation (\peq{16}) replaces equation (16) of [\pref{Dow2}].

The behaviours of $W^D$ and $W^N$ are now somewhat different to those plotted
in [\pref{Dow2}]. It is convenient to discuss the difference
$W_{\rm cap}(\si)-W_{\rm hemisphere}$. In the Dirichlet case this difference
is anti-symmetric about the point $\si=1$. It tends to $+\infty$ as the cap
shrinks ($\si\to0$) and tends to $-\infty$ as
the cap becomes the full sphere minus a point ($\si\to2$). In between, it has
a maximum ($\approx 8.9\times10^{-2}$) at
$\si=1+1/\sqrt2$ and a minimum at
$\si=1-1/\sqrt2$, corresponding to disc radii of $a=\sqrt2+1$ and
$a=\sqrt2-1$ respectively. It passes through zero at $\si=1$, the
hemisphere, and also at $\si\approx0.0425$ and $\si\approx1.9575$.

For Neumann conditions, the difference tends to $-\infty$ at both limits,
$\si\to0$ and $\si\to2$. It possesses a maximum ($\approx 3.1\times10^{-2}$)
at $\si=(5-\sqrt5)/2$ (\ie
at $a=5^{1/4}$) and goes through zero at $\si=1$ and at $\si\approx1.7195$.

The error in [\pref{Dow2}] has unfortunately propagated into a third paper
[\pref{Dow3}] where the
effective action on a spherical triangle (a sector of a spherical cap) was
found in terms of that on a sector of a disc using stereographic projection.

Equation (39) of [\pref{Dow3}] should read
$$
W_{\rm triangle}(\be,\si)=W_{\rm sector}(\be,\si)-
{\be\over12\pi}\big(\ln2+{1\over3}\si\big)
-{1\over24}\big({\pi\over\be}-{\be\over\pi}\big)\ln2 -{1\over8}\ln(2-\si)
$$
and likewise (40) becomes
$$\eqalign{
W_{\rm triangle}(\be,\si)=W_{\rm halflune}(\be)
&-{1\over48}\big(
3-{\pi\over\be}+{\be\over\pi}\big)\ln(2-\si)\cr
&-{1\over48}\big(3+{\pi\over\be}-{\be\over\pi}\big)\ln\si
+{\be(\si-1)\over6\pi}.\cr}
\eql{40}$$

Again, the contour plot of this quantity in Fig.4 in [\pref{Dow3}] does not
represent the correct behaviour. The saddle point occurs at the point
$\si\approx1,\,\be\approx60^\circ$.
\vskip 5truept

\noin{\bf{References}}
\vskip 1truept
\begin{putreferences}
\ref{Dow1}{J.S.Dowker and J.P.Schofield \jmp{31}{90}{808}.}
\ref{Dow2}{J.S.Dowker \cqg{11}{94}{557}.}
\ref{Dow3}{J.S.Dowker \jmp{35}{94}{4989}.}
\end{putreferences}
\bye